\documentclass[twocolumn,showpacs,prl,10pt]{revtex4}
\usepackage[english]{babel}
\usepackage{amsmath,amssymb} 
\usepackage{times}
\usepackage{epsfig} 

\begin{document} 

\title{Analysis of transport properties of iron pnictides:
  spin-fluctuation scenario}

\author{P. Prelov\v sek$^{1,2}$, I. Sega$^1$ and
T. Tohyama$^{3}$}
\affiliation{$^1$J.\ Stefan Institute, SI-1000 Ljubljana, Slovenia}
\affiliation{$^2$ Faculty of Mathematics and Physics, University of
Ljubljana, SI-1000 Ljubljana, Slovenia}
\affiliation{$^3$ Yukawa Institute for Theoretical Physics,
Kyoto University, Kyoto 606-8502, Japan}

\date{\today}

\begin{abstract}

We present a phenomenological theory of quasiparticle scattering and
transport relaxation in the normal state of iron pnictides based on
the simplified two-band model coupled via spin fluctuations. In
analogy with anomalous properties of cuprates it is shown that a large
and anomalous normal-state resistivity and thermopower can be
interpreted as the consequence of strong coupling to spin
fluctuations. The generalization to the superconducting phase is also
discussed.
\end{abstract}

\maketitle 

\section{I. Introduction}

Recently discovered superconducting (SC) iron pnictides
\cite{kami,ishi} are at present in the focus of experimental and
theoretical investigations in the solid state community. Besides high
SC $T_c$ the main motivation is the fact that a wide  class of
materials opens yet another view on the interplay of magnetism and SC
in metals. At present it seems that there is firm evidence that the
physics of novel materials is not in the class of strongly correlated
systems close to the Mott-Hubbard insulator as is the case in SC
cuprates \cite{imad}, investigated intensively in the last two
decades. On the other hand there are challenging similarities, in
particular in the anomalous transport properties \cite{lee,sefa,hess},
in the presumable unconventional type of SC \cite{hash,mazi}, as well
as in the importance of spin fluctuations apparently evidenced by the
NMR relaxation \cite{muku} and the onset of spin-density-wave (SDW)
order \cite{cruz}.

In this paper we put the emphasis on the transport properties of iron
pnictides (IP) and their understanding. Published experimental
results on the d.c. electrical resistivity $\rho(T)$ generally
reveal high values at $T\!>\!T_c$, comparable in magnitude to
underdoped cuprates \cite{ando}. At the same time, at elevated $T \sim
300$K the $T$-dependence is linear with large slope $d\rho/dT$, again
similar to the well known anomalous variation in cuprates. Since the
systematics of different compounds is still not fully understood we
concentrate here on the family emerging from the reference (undoped)
compound LnFeAsO (LFAO) where besides the original Ln=La a variety of
other lantanides Ln=Ce - Dy has been investigated so far. The electron
doping has been studied either by doping with F, i.e., in
LnFeAsO$_{1-x}$F$_x$ \cite{sefa,lee, hess} or via oxygen deficiency
LnFeAsO$_{1-y}$ \cite{eisa}.  The evidence so far is that different Ln
do show similar results, while the resistivity $\rho(T)$ reveals quite
systematic and universal change with doping $x$ \cite{hess} or $y$
\cite{eisa}. E.g., the resistivity \cite{hess,eisa} changes from that
of the SDW semimetal at $x,y<0.05$ with large $\rho(T\!>\!T_{SDW})$, a
property shared in particular by underdoped cuprates, over the
intermediate (optimally doped) regime $x,y \sim 0.1$ with a quite
pronounced linear law $\rho \propto T$, into the overdoped regime with
more Fermi-liquid (FL) $T^2$ behavior for $y>0.2$. 

Also the thermopower $S(T)$ is far from the FL behavior
$S \propto T$ \cite{lee,sefa}, reaching at $T\!>\!T_c$ values
characteristic for nondegenerate electrons, i.e., $|S| \sim
s_0=k_B/e_0=86 \mu$V/K again being a remarkable property of
underdoped cuprates \cite{coop}. Similar message is emerging from
strongly $T$-dependent Hall constant $R_H(T)$ \cite{koha}. So far,
there are very few data on dynamical transport properties, nevertheless
optical conductivity in the same system \cite{qazi} seems to support
non-Drude-like relaxation with large and $\omega$-dependent transport
relaxation rate $\tau^{-1} \sim \omega$.

A detailed analysis of transport data on IP seems to be still
premature due to mostly polycrystalline samples studied so far as well
as due to the lack of doping systematics in the electron subsystem.
Nevertheless in the following we argue that the similarity to cuprates
arises from the strong coupling to spin fluctuations and the
non-FL-like behavior following the marginal FL (MFL) scenario
\cite{varm}. More specific origin of spin fluctuations and the
spin-fermion coupling could be, however, quite different, e.g., due to
the importance of the Hund's rule coupling $J_H$, also shown to lead
to large $\rho(T)$ \cite{haul}. 
Also, emerging novel results on single-crystal transport obtained
mostly by the electron doping the reference compound BaFe$_2$As$_2$
\cite{naka,fang} require a quantitative reconsideration, not followed in
this work in detail, due to observed higher mobilities. Nevertheless,
several main characteristics as the linear $\rho(T)$ law, large and
anomalous thermopower $S(T)$, $T$-dependent Hall constant $R_H(T)$ again
confirm our basic assumptions given below.

\section{II. A simplified model}

A microscopic model for relevant electrons in IP appears to be quite
complicated and still debated at present \cite{mazi}. While it has
been argued that a (minimal) two-band model already contains the
essential low energy physics in these materials \cite{ragh,chen,kors},
the inadequacy of certain approximations to the effective low-energy
band structure has been recently criticized \cite{gras}. Nevertheless,
for the purpose of our qualitative analysis we employ the model with
two bands, one electron-like and another hole-like, coupled, however,
through spin fluctuations introduced phenomenologically \cite{chub}
and treated within the lowest-order pertubation theory.  It is evident that
such a simplified model is not enough to describe calculated band structure
\cite{sing} as well as the one observed via angle resolved photoemission
\cite{liu} or via the de Haas-van Alphen effect  \cite{cold}, revealing (at
least) four pockets at the Fermi surface in the actual Brillouin
zone. However, in the simplest approach to  spin-fluctuation mechanism it is
essential that SDW-type spin fluctuation inter-couple electron and hole
bands, while the coupling among electron (hole) bands themselves should be
less important. 

In contrast to IP, a vast experimental evidence in the last two decades
shows that cuprates can be well modeled with a single-band model
\cite{imad}, e.g., the simple 2D Hubbard model or the $t$-$J$ model
\cite{jakl} being, however, in the regime of strong correlations which can
be only approximately described with a coupled spin-fermion model
\cite{chub}. Starting from the latter level assuming the MFL behavior of
spin fluctuations and a strong coupling to electrons yield in cuprates
anomalous quasiparticle (QP) damping \cite{pr} and transport relaxation
\cite{zeml} as well as an  unconventional SC \cite{pr}.

We use a simplified model for IP describing the 2D system with one
electron (e) band and the other hole (h) band crossing the Fermi
surface \cite{mazi,ragh,chen}. I.e., in the (folded) Brillouin zone
\cite{kors,mazi} the h-like and e-like pockets are at $k \sim 0$, and
${\bf k} \sim {\bf Q} =(\pi,\pi)$, respectively. Within this effective
model bands are coupled only via spin fluctuations, leading to
\begin{eqnarray}
\hspace{-0.15in}H_{ef}\hspace{-0.15in}&&=\!-\!\sum_{{\bf k},s}
\bigl(\zeta^e_{\bf k} c^\dagger_{{\bf k}s} c_{{\bf k}s} + \zeta^h_{\bf
  k} d^\dagger_{{\bf k}s} d_{{\bf k}s} \bigr) + \nonumber \\ &&
\frac{1}{\sqrt{N}} \sum_{{\bf kq},ss^\prime} m_{\bf kq} {\bf S}_{\bf
  q}\cdot {\bf \sigma}_{ss\prime}( c^\dagger_{{\bf k}-{\bf q},s}
d_{{\bf k}s'}+ d^\dagger_{{\bf k}-{\bf q},s} c_{{\bf
    k}s'}), \label{mod}
\end{eqnarray}
and $c_{\bf k},d_{\bf k}$, ($\zeta^e$, $\zeta^h$) refer to
electrons in e-like and h-like bands, respectively.
We consider the corresponding Green's functions for e- and h-electrons
\begin{equation}
G_{\bf k}^{\sigma}(\omega)= [\omega^+ - \epsilon^{\sigma}_{\bf k} 
- \Sigma^{\sigma}_{\bf k}(\omega)]^{-1}, \label{gf}
\end{equation}
where $\epsilon^{\sigma}_{\bf k} =\zeta^{\sigma}_{\bf k} -\mu$, and
$\sigma=e,h$ ($\bar\sigma=h,e$).

\section{III. Quasiparticle damping}

Within the lowest order perturbation 
theory \cite{pr,plak} the self energies can be expressed as
\begin{eqnarray}
\Sigma^{\sigma}_{\bf k}(\omega)\!\!\!\!&&=3 \sum_{\bf q} m^2_{\bf k q} \int\!
\!\int \frac{d\omega_1 d\omega_2}{\pi} g_{12} \frac{A^{\bar\sigma}_{{\bf k}-{\bf
q}} (\omega_1) \chi''_{\bf q}(\omega_2)}
{\omega-\omega_1-\omega_2 }, \nonumber \\
g_{12}\!\!\!\!&&\equiv g(\omega_1,\omega_2)= \frac{1}{2}\bigl [{\rm
th}\frac{\beta\omega_1}{2}+{\rm cth}\frac{\beta\omega_2}{2} \bigr], 
\label{sig}  
\end{eqnarray}
where $\chi_{\bf q}(\omega)$ is the dynamical spin susceptibility.

To proceed we make several simplifications, which are expected to be a
reasonable starting point for a qualitative analysis of transport
quantities in IP. Spin response close to the antiferromagnetic (SDW)
instability \cite{cruz} centered at ${\bf q} \sim {\bf Q} = (\pi,\pi)$
we assume broad enough relative to h/e pockets to replace $\chi_{\bf
  q}(\omega) \sim \chi_{\bf Q}(\omega) =\tilde \chi(\omega)$. In this
case $\Sigma^e_{\bf k}\sim \Sigma^e_{\bf Q}= \Sigma^e$ and
$\Sigma^h_{\bf k}\sim \Sigma^h_{\bf 0}=\Sigma^h $, and the QP damping
is $\Gamma^{\sigma}(\omega)=-\mathrm{Im}\Sigma^{\sigma}(\omega)$.

It should be however pointed out that the momentum dependence of the self
energy could be important or even crucial. The latter would be the case if,
e.g., the (accurate) nesting would play a role. Since anomalous behavior of
IP seems to be quite robust, e.g., linear resistivity over a broad $T$ and
doping range, large thermopower $S(T)$ as well as high $T_c$, we believe
that this is not the case so that momentum dependence along the Fermi
pockets is not crucial also requiring $\chi_{\bf q}(\omega)$ response not
too narrow in ${\bf q}$.

With the above simplifications we get for the QP damping
\begin{equation}
\Gamma^{\sigma}(\omega)=
\frac{3}{2}  \lambda\int d\omega' g(\omega-\omega',\omega') {\cal
  N}^{\bar\sigma}(\omega-\omega') \tilde \chi''(\omega'), \label{sigl}
\end{equation}
where $m_{\bf Q,Q} = m_{\bf 0,Q} =\bar m$, $\lambda=\bar m^2$, and
${\cal N}^{\sigma}(\omega)$ are the $\sigma$-band density of states
(DOS). We also get
\begin{equation}
 A_{\bf k}(\omega)=-\mathrm {Im}[\Omega(\omega) - \epsilon_{\bf k} 
+i\Gamma(\omega)]^{-1}, 
\end{equation}
where $\Omega(\omega)=\omega-\mathrm{Re}\Sigma(\omega)
=\omega Z^{-1}(\omega)$ defines the QP weight $Z(\omega)$. 

In analogy with cuprates, large and non-FL-like linear
resistivity $\rho(T)\propto T$ and a presumable transport relaxation
rate $\tau^{-1} \propto \omega$ in particular, require also a MFL
behavior \cite{varm} of the spin fluctuation input $\chi(\omega)$. In our
analysis MFL-type spin fluctuations are a phenomenological assumption which
still has to be confirmed by specific experiments as, e.g., the inelastic
neutron scattering. It is also fair to admit that a theoretical
understanding of such spin fluctuations in IP is not yet available, whereby
even for cuprates there is not yet an agreement on the explanation for the
MFL physics. Still, as for cuprates there are no accepted
alternative scenarios for the non-FL behavior of QP damping and transport
quantities.

Hence, we employ the relation \cite{jakl}
\begin{equation}
\tilde \chi''(\omega)=\pi \bar C(\omega) {\rm th}(\beta
  \omega/2),
\label{chi}
\end{equation}
where $\bar C(\omega)=C(\omega)+C(-\omega)$ and $C(\omega)$ represents
the dynamical spin correlation function. In analogy with the MFL
scenario \cite{varm} well established for cuprates we use a
$T$-independent ansatz $C(\omega) \sim C_0, ~\omega<\omega_0
\gg T$ \cite{jakl,zeml}. It should be noted that such an ansatz by
construction satisfies the $T$-independent sum rule \cite{psb} given by
\begin{equation}
\frac{1}{\pi} \int_0^\infty {\rm cth}\frac{\beta \omega}{2}\tilde
\chi''(\omega) d\omega=\int_0^\infty \bar C(\omega) d\omega=\langle
S^z_i S^z_i \rangle,
\label{sumrule}
\end{equation}
In underdoped cuprates the MFL behavior emerging from the above
assumptions is qualitatively rather well established both
experimentally \cite{kast,imad} and from model calculations
\cite{jakl}, although other forms close to Eq.(\ref{chi}) have also
been proposed \cite{kast}.  The origin of a non-FL behavior seems to 
emerge from the fact that due to the localized character of spins the
low-$\omega$ spin-fluctuations exhaust the sum rule \cite{psb}. This
is much less evident for IP but still appears to be a prerequisite for the
MFL-like behavior of transport quantities as described furtheron.

Assuming in the relevant regime ${\cal N}(\omega) \sim {\cal
  N}$ (being to lowest order unchanged even if the QP weight $Z<1$) we
get
\begin{eqnarray}
\Gamma^\sigma(\omega)= \gamma^\sigma\omega{\rm
coth}(\beta\omega/2),
\,\, \gamma^\sigma=(3\pi/2)\lambda C_0\cal{N}^{\bar\sigma},
\label{gmfl}
\end{eqnarray}
i.e., approximately
\begin{equation}
\Gamma^\sigma(\omega)\cong\gamma^\sigma\mathrm{max} (|\omega|,2T) .
\end{equation}
However, for further interpretation, in particular of the Seebeck
coefficient $S(T)$, it seems essential that the DOS be nonsymmetric
around $\omega \sim 0$. E.g., a possible assumption is ${\cal N} (\omega
\gtrless 0) \sim {\cal N}_\pm$, where the DOS can differ 
for $\omega \gtrless 0$ and consequently $\gamma^{\sigma} \to
\gamma^{\sigma}_{\pm}$. While the asymmetry of the relaxation rate seems to
be the only viable explanation for the large $S(T)$, assumptions are
expected to emerge from a more detailed analysis of ${\cal N}(\omega \sim
0)$ in a doped semimetal and corresponding scattering rates for $\omega \pm
0$. E.g., due to very shallow hole bands at $\omega>0$ for electron
scattering in e-doped IP \cite{liu,seki} as  considered below one is
effectively dealing with different ${\cal N}_\pm$.

\section{IV. Transport quantities}

Turning to transport properties, we
first consider optical conductivity $\sigma(\omega)$ which we assume
isotropic within the Fe-As, i.e., easy plane. Within the linear
response (neglecting vertex corrections) $\sigma(\omega)$ can be
expressed as \cite{maha}
\begin{eqnarray}
&&\sigma(\omega)=\frac{2 \pi e_0^2}{N\omega} \sum_{{\bf k},\sigma} (v_{{\bf
    k}\sigma}^x)^2 \times\nonumber\\&&\times\int d\omega' [f(\omega'-
    \omega)-f(\omega')] A^\sigma_{\bf k}(\omega')A^\sigma_{\bf
    k}(\omega'-\omega), \label{sigom}
\end{eqnarray}
where $v_{{\bf k}\sigma}^x$ are the corresponding band velocities.  In
the following we consider e-doped IP, therefore for simplicity we take as
dominant the e-pocket contribution. Everywhere refering to the e-band and
defining the function
\begin{equation}
\Phi(\epsilon)=\frac{2 e_0^2}{N} \sum_{\bf k}(v^{x}_{\bf k})^2
\delta(\epsilon-\epsilon_{\bf k}), \label{phi}
\end{equation} 
we get for slowly varying $\Phi(\epsilon) \sim \Phi_0$
\begin{equation}
\sigma(\omega)=\Phi_0\frac{1-e^{-\beta\omega}}{\omega}
\!\int d \omega' \frac{f(-\omega')f(\omega'-\omega)F_{12}} {\bar
  \Omega_{12}^2 + F^2_{12}}, \label{sigf}
\end{equation}
where $F_{12}=\Gamma (\omega')+ \Gamma
(\omega'-\omega)$ and $\bar\Omega_{12}=\Omega
(\omega')-\Omega (\omega'-\omega) \sim Z^{-1} \omega$.

For $\sigma(\omega \gg T)$ one gets from Eq.(\ref{sigf}) the extended
Drude form
\begin{equation}
\sigma(\omega \gg T)= \tilde \Phi \frac{\Gamma_{tr}(\omega)}{\omega^2+ 
\Gamma_{tr}^2(\omega)},
\label{edru}
\end{equation}
and $\tilde \Phi=Z \Phi_0 \sim n_e e_0^2/m_e^*$, with $m_e^*$ the QP
mass in the e-pocket. Assuming further the MFL form for
$\Gamma(\omega)$, Eq.(\ref{gmfl}), the effective transport relaxation
rate $\Gamma_{tr}(\omega)= \tilde \gamma_a \omega$ where $\tilde
\gamma_a=Z (\gamma_+ +\gamma_-)/2$.

For the d.c. conductivity Eq.(\ref{sigf}) reduces to \cite{maha,pals}
\begin{equation}
\sigma(0)= \tilde \Phi \int d\omega\biggl(-\frac{\partial f}{\partial
  \omega}\biggr) \frac{1} {\Gamma_{tr}(\omega)}, \label{sig0}
\end{equation}
which, on assuming constant $\tilde \Phi$ and MFL-type $\Gamma$,
Eq.(\ref{gmfl}), immediately yields linear-in-$T$ resistivity
\begin{equation}
\rho=\frac{T}{A_0}=\frac{T}{\tilde \Phi \tilde A_0}. \label{rho0}
\end{equation}
Within the same local approximation for 
$\Sigma_{\bf k}(\omega)=\Sigma(\omega)$ also the Seebeck coefficient $S$ can
be expressed as \cite{pals}
\begin{equation}
S=- w s_0, \qquad w=\tilde A_1/\tilde A_0, \label{s0}
\end{equation}
where
\begin{equation}
\tilde A_n= T \int d\omega\biggl(-\frac{\partial f}{\partial \omega}\biggr)
\frac{(\beta \omega)^n}{2 Z\Gamma(\omega)}. \label{an}
\end{equation}
Under the MFL assumption for $\Gamma(\omega)$, all $\tilde A_n$ are
$T$-independent and in contrast to the FL behavior $S \propto T$ 
one gets a $T$-independent $S \sim~$const.

In the symmetric case $\gamma_-=\gamma_+=\gamma$ and by
Eqs.(\ref{gmfl}),(\ref{an}) 
\begin{equation}
\tilde A_0= 0.21/\tilde \gamma,
\end{equation}
while $A_1=0$ identically and therefore $S(T)=0$. It is thus evident
that {\it a pronounced asymmetry} in $\Gamma(\omega)$ is needed to explain
large $S(T)$ in IP, as discussed later. One situation possibly
relevant for e-doped IP is that $\gamma_- \gg \gamma_+$. This can
happen if upon electron doping the h-pocket states diminish
substantially at $\omega>0$ as a source of scattering, leading to the
reduction of $\gamma_+$. In such a limiting case we get
\begin{equation}
\tilde A_0=0.10/\tilde \gamma_+,\qquad w\sim 1.2~.
\label{asy}
\end{equation}

\section{V. Analysis of experimental results}

So far, most information for
the normal(N)-state transport in IP is available for the
d.c. resistivity $\rho(T)$. We analyse here only the data for electron
doped LFAO compounds. The whole range of $x$ for F-doped LFAO has been
measured recently \cite{hess}, while $S(T)$ as well as Hall
coefficient $R_H(T)$ have been measured also for $x=0.11$
\cite{sefa,lee} and $x=0.05$ \cite{koha}. It should be reminded, however,
that all data so far are for polycrystals while the theory is
done for the transport oriented along the easy plane, so that measured
$\rho(T)$ should be at least scaled by some factor $\xi<1$ to get the
relevant planar resistivity considered here.

First, it is evident that inverse mobilities are larger in IP as compared
to cuprates. Thus in LaSr$_x$CuO$_{1-x}$ at $T=300$K and for doping $x=0.03
- 0.1$ 
\cite{ando},  
\begin{equation}
\mu^{-1} = n_h e_0 \rho_{ab} = (0.3 - 0.15) \mathrm{Vs/cm}^2.
\end{equation}
Analogous results for IP compounds depend on the density  $\bar x$ of
carriers/formula unit, where $n_e=\bar x/V_0$ and $V_0$ is
the volume of a formula unit. Even undoped $x=0$ IP have finite but
small $n_e^0$, i.e., $x_0>0$ and it is plausible that $\bar x=x_0 +x >
x$. From data at $T=300$K
\cite{hess} we get
\begin{equation}
\mu^{-1}= n_e e_0 \rho(T) \sim (5.6 - 8.1) \bar x~\mathrm{Vs/cm}^2,
\end{equation}
for $x=0.05 - 0.2$, respectively. Similar results are obtained for
$x=0.11$ \cite{sefa,lee} $\mu^{-1}= 5.8 \bar x $Vs/cm$^2$, and for
$x=0.05$ \cite{koha} $\mu^{-1}= 8.0 \bar x $Vs/cm$^2$.  It is
evident that such $\mu^{-1}$ are even higher than in low-doped
cuprates \cite{ando}.

Existing data for $\rho(T)$ in LFAO show a large slope at higher $T
\gg T_c$, also with a larger onset $\rho(T \sim T_c)$ at $x=0.05$, while
for $x>0.1$ $\rho(T<T^* \sim 100K)$ becomes nonlinear and more
FL-like.  To estimate $\tilde A_0$ and consequently $\tilde \gamma$ we use
the slope at $T < 300$K, i.e., $\tilde A_0^{-1}=\tilde\Phi d\rho(T)/dT$ and
we get for a wide range of $x$ \cite{hess}
\begin{equation}
\tilde A_0 \sim \alpha\nu/\bar x, \qquad \alpha \sim
3.8\cdot 10^{-3}, \qquad \tilde \gamma \sim 56 \bar x/\nu, \label{a0fit}
\end{equation}
where $\nu=m/m^*$.  Similar values are obtained analysing other data
for F-doped LFAO: for $x=0.11$ \cite{lee} and for $x=0.05$ we get
$\alpha \sim 4.0\cdot 10^{-3}$ \cite{koha}. Not much different is the
development of $\rho(T)$ for the oxygen deficient LFAO \cite{eisa} where   
$y<0.03$ compounds reveal a non-SC state with a SDW transition $T_{SDW}>0$
while for $y>0.1$ again  $\rho(T)$ is nicely linear in $T$ with $\alpha \sim
0.006$.  Assuming, e.g., values from band structure calculations $\nu
\sim 2$ and $\bar x \sim 0.1$ we arrive at very large $\tilde \gamma
\sim 2.8$. Such value $\tilde \gamma$ is most likely an overestimate due
to too large $\rho$ (all cited measurements are for polycrystals) and possibly
due to relatively small QP mass enhancement $\nu \sim 2$. 

So far, there are only few data for optical conductivity
$\sigma(\omega)$. In the LFAO with As replaced by P $\sigma(\omega)$
was measured and analysed \cite{qazi} using the extended Drude fit
yielding for $\omega<1000$cm$^{-1}$ anomalous $\tau^{-1}=\Gamma_{tr}
\sim \omega$, i.e., $\tilde \gamma \sim 1$ qualitatively consistent
with the above estimates.

Experimental results for $S(T)$ for LFAO yield typically e-like $S<0$
with strong $T$-dependence with the maximum values $S \sim -
s_0$ at $T \sim 100$K. Assuming $\Delta \tilde\gamma = \tilde \gamma_-
- \tilde \gamma_+ > \tilde \gamma_+$ we get from Eq.(\ref{an}) 
\begin{equation}
w \sim\Delta \tilde \gamma/(12 \tilde \gamma_a)\nonumber
\end{equation}
 whereas in the extreme
asymmetric limit, Eq.(\ref{asy}), we recover $w=1.2$.

\section{VI. Superconductivity}

Let us finally comment on the relation of the above analysis to the
treatment of the SC phase within the assumption of the spin-fluctuation
induced pairing. We follow here closely the treatment of SC within the
effective spin-fermion model \cite{plak,pr} as derived from the microscopic
strong-correlation model (planar $t$-$J$ model) relevant for cuprates. Since
our paper is not focused on the question of SC in IP, the aim is to connect
parameters entering transport quantities to those determining the SC gaps
and consequently $T_c$.

For the discussion of SC equations can be generalized with Green's
functions and selfenergies being $2\times 2$ matrices \cite{plak}
whereby we (again) neglect e/h interband terms
\begin{equation}
{\bf G}^{\sigma}_{\bf k}(\omega)= [\omega \tau_0 - \epsilon^{\sigma}_{\bf k} 
\tau_1- {\bf \Sigma}^{\sigma}_{\bf k}(\omega)]^{-1}. \label{gsc}
\end{equation}
In analogy to the normal state the lowest-order approximation for
the self-energy can be written as \cite{pr,plak}
\begin{equation}
{\bf \Sigma}^{\sigma}_{\bf k}(i \omega_n)=\frac{-3}{N\beta}
\sum_{{\bf q},m} m^2_{\bf kq} {\bf G}^{\sigma}_{{\bf k}-{\bf
q}}(i \omega_m) \chi_{\bf q}(i \omega_n-i \omega_m) ,\label{sigma}
\end{equation}
where $i \omega_n=i \pi(2n+1)/\beta$. Again neglecting the ${\bf k}$
dependence within each band, i.e., ${\bf \Sigma}^{\sigma}_{\bf k}={\bf
  \Sigma}^{\sigma}$ one gets from Eq.(\ref{sig}) a nonzero gap
$\Delta^{\sigma} \sim Z^{\sigma} \Sigma_{12}^{\sigma}(0)$, and
\begin{equation}
\Delta^{\sigma}=\frac{-3\lambda}{N}\sum_{\bf q} \chi^0_{\bf q}
C^{\bar\sigma}_{\bf q} \frac{Z^{e}
Z^{h} \Delta^{\bar\sigma}}{2 E^{\bar\sigma}_{\bf q}} \mathrm{th}\frac{\beta
E^{\bar\sigma}_{\bf q}}{2}, \label{gapeh}
\end{equation}
where $(E^e_{\bf q})^2=\varepsilon^2_{\bf q}+(\Delta^e)^2$,
$(E^h_{\bf q})^2=\varepsilon^2_{{\bf q}-{\bf Q}}+(\Delta^h)^2$
and $C^{\bar\sigma}_{\bf q}=I^{\bar\sigma}_{\bf q}(i\omega_n \sim 0)/\tilde 
I^{\bar\sigma}_{\bf q}$ plays the role of the cutoff function with
\begin{equation}
I^{\sigma}_{\bf q}(i\omega_n)=\frac{1}{\beta \chi^0} \sum_{m} \chi(i\omega_n-
i\omega_m) \frac{1}{\omega_m^2 + (E^{\sigma}_{\bf q})^2}, \label{iq}
\end{equation}
and $\tilde I^{\sigma}_{\bf q}= \mathrm{th}(\beta E^{\sigma}_{\bf q}/2)/
(2 E_{\bf q})$. 

Finally, at $T=0$ Eqs.(\ref{gapeh}), (\ref{iq}) reduce to
\begin{equation}
\Delta^{\sigma}=-\frac{3}{2}\lambda \chi^0 Z^{\sigma} \Delta^{\bar\sigma}
\int_{\omega_c^{\bar\sigma}}^{\omega_c^{\bar\sigma}} d\varepsilon
\frac{{\cal N}^{\bar\sigma}(\varepsilon)}
{\sqrt{\varepsilon^2 + (\Delta^{\bar\sigma})^2}} , \label{gapt0}
\end{equation}
where $\omega_c^{\sigma}$ are effective cutoffs. It is evident from
Eq.(\ref{gapt0}) that SC is of the $s_\pm$ - type \cite{mazi}, that is
$\Delta^h = - \eta \Delta^e$, $\eta > 0$. Assuming also ${\cal
  N}^{\sigma}(\varepsilon) \sim {\cal N}^{\sigma}$ we get from
Eqs.(\ref{gapt0}),(\ref{gmfl}),
\begin{equation}
1= \tilde \gamma^e \tilde \gamma^h 
B^2 \ln \frac{\omega_c^h}{\Delta^h}\ln \frac{\omega_c^e}{\Delta^e}, 
\label{bcs1}
\end{equation}
with $B =(4/\pi)\ln (\omega_0/2T)$ connecting qualitatively the N-state
transport parameters with the gap equation. Clearly the SC Eliashberg
equations are treated in a simplified manner \cite{dolg} in order to get the
familiar BCS-type form. Still the message is quite clear: spin-fluctuation
mediated interaction gives naturally the $s_\pm$ - type SC pairing
consistent with other approaches \cite{mazi}. Parameters entering
Eq.~(\ref{bcs1}) are besides $B \sim {\cal O}(1)$ (depending on the form
of the  spin-fluctuation spectra) the cutoffs $\omega_c^\sigma$ determined
by the characteristic spin fluctuation frequency and $\tilde
\gamma^{\sigma}$. The latter have clearly the strongest influence and
according to our estimates from transport, Eq.(\ref{a0fit}),
$\tilde\gamma^{\sigma} >1 $ are large requiring the strong coupling approach both
for the N and for the SC state. As discussed in the next section on the
basis of emerging single-crystal results, smaller resistivities
$\rho(T)$ are reported and consequently $\tilde \gamma<1$. Still $\tilde
A_0$ and the coupling remain at least moderate and in the same range as in
optimum-doped cuprates, giving support for, or at least not contradicting,
the notion for an SC pairing mechanism and its strength common to both  IP and cuprates. 

\section{VII. Discussion}

We have presented a theory based on the
spin-fluctuation induced coupling between the e- and h-bands in IP
with the motivation to explain their anomalous N-state transport
properties. Existing experimental data on polycrystalline samples indicate
that the QP damping and transport relaxation rates are even higher than in
underdoped cuprates. It seems rather unlikely that quantitatively similar
results should obtain for single crystals, as evidenced quite recently by
measurements on, e.g., BaFe$_{\rm 2}$(As$_{\rm 1-x}$P$_{\rm x}$)$_{\rm 2}$,
a material from the 122 family of IP \cite{kasa} or BaFe$_{\rm
2-x}$Co$_{\rm x}$As$_{\rm _2}$ \cite{fang} where $\tilde\gamma$ and $1/\mu$
are substantially reduced with respect to values in (polycrystalline) LFAO
samples of comparable doping. However, the marked linearity of $\rho(T)\sim
T/A_0+{\rm const.}$ observed in BaFe$_{\rm 2}$(As$_{\rm 1-x}$P$_{\rm
x}$)$_{\rm 2}$ over most of the  doping region with nonzero $T_c$ testifies
to the non-FL behavior, similar to cuprates. 
Likewise the evolution with doping in LFAO compounds is quite analogous,
from a near insulator in an undoped substance to a FL-like behavior in the
overdoped IP. Observe, however, that for the NdFeAsO$_{\rm 1-x}$F$_{\rm x}$
compound there is not much difference between polycrystalline and single
crystal data concerning $\rho(T)$ and, e.g.,  $\alpha$, Eq.(\ref{a0fit}).
Thus, for $x=0.18$ single crystal sample \cite{cheng} a rough estimate for
$T\sim 200 {\rm K} - 300 {\rm K}$ yields $\alpha\sim 0.023$ whereas in a
polycrystalline sample $x=0.11$ \cite{ren} $\alpha\sim 0.009$, resulting in
$\tilde\gamma\sim 0.5$ and $1.1$, respectively, i.e., again comparable to
cuprates. Thus a more systematic study of the transport and optical
properties in single crystal compounds of the separate families of
oxypnictides is needed to settle this issue, particularly in view of the
recent analysis of the competition between the linear and quadratic in $T$
contributions to $\rho(T)$, where the former is seen to dominate $\rho(T)$
in samples with the highest $T_c$ for the  compounds there considered \cite{doir,ishi}.

Certain conclusions emerging from the above analysis still seem to be hard
to avoid: a) the coupling to spin fluctuations is apparently substantial so
that the QP damping is large with the QP overdamped in the low-doping regime,
b) the effect of spin fluctuations on the N-state transport and on the
SC pairing likewise appears to be strong implying pronounced spin fluctuations in the
low-frequency window, both properties shared by underdoped and optimally doped
cuprates as well, c) if estimates emerging from experiments are correct the
strength of the coupling could be beyond the applicability of the
lowest-order perturbation theory employed here, d) the behavior of IP even
at modest $T>T_c$ is non-FL-like as clearly evidenced by large $S(T)$ as
well as the $T$-dependence of $R_H(T)$, whereby the common features with
cuprates stem from the large spin-fermion coupling and not from the
Mott-Hubbard physics. However, rather scarce experimental evidence for
low-energy spin fluctuations requires some caution and additional  efforts
to pin down the proper ingredients for a viable theory of IP.

\section{Acknowledgements}

Authors acknowledge fruitful discussions with C. Hess and the access
to their unpublished data, and the financial support of MHEST and JPSJ
under the Slovenia-Japan Research Cooperative
Program. T.T. acknowledges the support of the TRIP project.

\end{document}